\documentstyle[manuscript,aps,version2,epsf]{revtex}
\begin{document}
\draft
%\preprint{OU-HEP XXX}
\title
\bf  Continuum Theory with Memory for Avalanches \\
in Self-Organized Criticality
\endtitle

\author{Maya Paczuski$^1$ and Stefan Boettcher$^{2}$}
\instit
$^1$Department of Physics, Brookhaven National Laboratory,
Upton, NY 11973\\
$^2$Department of Physics and Astronomy, The
University of Oklahoma, Norman, OK 73019-0225
\endinstit
%\receipt{\today}
\medskip
\centerline{\today}
\medskip

\abstract 
 The propagator for the activity in a broad class of
self-organized critical models obeys an
imaginary-time Schr\"odinger equation with a nonlocal,
history-dependent potential representing memory.  Consequently, the
probability for an avalanche to spread beyond a distance $r$ in time
$t$ has an anomalous tail $\exp{[-C\,x^{1/(D-1)}]}$ for $x=r^D/t \gg
1$ and $D>2$, indicative of glassy dynamics.  The theory is
verified for an exactly solvable model, where $D=4$ and
$C=3/4$, and for the Bak-Sneppen model where it is tested numerically.
\endabstract

\pacs{PACS number(s): 64.60.Lx, 64.70.Pf, 87.10.+e}

At the simplest level, the dynamics of large, extended systems may be
either fast and chaotic, or slow and possibly frozen.  In the first
case information about the current microscopic state decreases rapidly
in time; in the second the system acquires long-term memory.  As some
tuning parameter is varied, there could generally be a critical
transition between these two phases where the dynamics takes place
intermittently in terms of avalanches.  Possible examples include
depinning transitions \cite{scaling}, dynamic glass transitions
\cite{dgl}, and certain transitions in cellular automata
\cite{macready}.  Many slowly driven, dissipative systems may
naturally operate at this transition; the criticality is
self-organized (SOC) \cite{btw}.  Such systems evolve in terms of
scale-free avalanches sweeping the system.  This picture can be
verified in a broad class of SOC models \cite{scaling} including
interface \cite{sneppen} and flux-line depinning \cite{zaitsev}, and
certain aspects of biological evolution \cite{BS,BoPa}.  These models
represent constant-velocity freezing transitions where the velocity $v
\rightarrow 0$ of motion is the order parameter.  The 
intermittent, punctuated equilibrium behavior at the phase transition
$(v=0)$ combines features of frozen systems with those of chaotic
ones.  Such systems can remember the past because of the long periods
of stasis allowing them to preserve what has been learned through
history; they can evolve rapidly and adapt because of the intermittent
bursts of activity, which span all length scales up to the system
size.

The freezing transition is an attractor for the dynamics when 
the site which is updated at each instant has the
current, globally extremal value of some force.  The extremal update
causes that site as well as neighboring sites to get new forces
according to specific model-dependent rules.  In the stationary critical state,
the pattern of activity is correlated and can be described in terms of
scale free avalanches.  The avalanche propagator $G(r,t)$ is the conditional
probability to find the updating activity at $r$ at time $t$ given a surviving
avalanche which started at the origin at time $t=0$.  Knowledge of the
propagator as well as the distribution of avalanche durations contains
all of the information on the macroscopic behavior of the system.  Any
universal quantities which may exists would appear in the asymptotic
long length and long time limits of these functions.

Our main result is that the asymptotic behavior of the propagator, $G$, is
governed by an imaginary-time Schr\"odinger equation
with a nonlocal, history-dependent potential
\begin{equation}
{\partial G(r,t) \over \partial t} = \nabla_r^2 G(r,t)
+ \int_0^t V(t-t')G(r,t')dt' \quad .
\label{e1}
\end{equation}
As the system builds up to the critical state the potential 
becomes scale free, 
\begin{eqnarray}
V(t)\sim t^{-\alpha},\quad (t\to\infty).
\label{e2}
\end{eqnarray}
Such dynamics always leads to subdiffusive spreading of the avalanche,
with an avalanche dimension $D>2$, where $D=2/(\alpha-1)$.
The equation governing the asymptotic behavior of
the propagator adheres to an anomalous non-Gaussian diffusion form,
\begin{eqnarray}
G(r,t)\sim A(r,t) \exp\left[-C\left({r^D\over
t}\right)^{1\over D-1}\right] \quad (r^D\gg t\gg 1), 
\label{n4}
\end{eqnarray}
where $C$ is a constant and $A(r,t)$ contains only powers of $r$ and
$t$.  For the Bak-Sneppen evolution model we have verified this
prediction numerically as shown in Figs.~1 and ~2. In a similar model,
the multi-trait evolution model \cite{BoPa}, Eqs.~(\ref{e1}-\ref{n4})
can be explicitly derived from microscopic rules.  In this case, the
nonlocal potential $V(t)$ is obtained exactly; it is the distribution
of avalanche durations.  The resulting equation for the propagator is
solved exactly with $\alpha = 3/2$ to give Eq.~(\ref{n4}) with $D=4$
and $C=3/4$.    

Similar dynamical equations with memory
describing aging of glasses and spin glasses within mode-coupling
theory have been introduced recently \cite{dgl,cule}.  This similarity
suggests a possible connection between ergodicity breaking in glasses
and the punctuated equilibrium behavior of self-organized critical
systems.  Perhaps, like SOC, the dynamics of glasses is at the edge
separating frozen and chaotic phases, although previous results
testing this view have been mixed \cite{tangbak,jacobs}.

The simplest extremal model is the Bak-Sneppen evolution model defined
as follows \cite{BS}: Put a random number at each site on a $d$
dimensional lattice.  At each time step, replace the smallest random
number in the system and its $2d$ neighbors with new random numbers
drawn independently from a flat distribution between 0 and 1.  After a
long transient the model self-organizes to a critical state where
almost all sites have values above a threshold $\lambda_c$.  An
avalanche of duration $t$ begins at time $0$ when the minimal random
number in the system equals $\lambda_c$ and ends at the first instant
in time, $t$, when the minimal random number reaches $\lambda_c$
again.  The sequence of values below $\lambda_c$ form bursts which
span all spatial and temporal extents.  Reducing the threshold level
$\lambda < \lambda_c$ divides the critical avalanches into smaller
$\lambda$ subavalanches (with a cutoff) which form a hierarchy
\cite{scaling}.  For any $\lambda < \lambda_0 < \lambda_c$ the
$\lambda_o$ subavalanche must persist longer than the $\lambda$
sub-avalanche that it contains.  In all extremal models, the avalanche
hierarchy is characterized by certain power laws \cite{scaling}.
Thus, a mechanism for long-term memory is apparent.

We propose that the
propagator $G$ for the avalanches in these models obeys
Eqs.~(\ref{e1},\ref{e2}) for $r^D\gg t\gg 1$.  Current activity has
some probability to activate neighboring sites at subsequent nearby
times.  In an isotropic system, this by itself would give ordinary
diffusion as represented by Eq. (\ref{e1}) with $V=0$.  The memory
term with nonzero $V$ explicitly breaks invariance with respect to
shifts in time.  The history-dependent potential $V(t)$ is model
dependent. This potential accounts for the avalanche hierarchy
described above.  We conjecture that in all extremal SOC models $V(t)$ is
a power law for large arguments $t$; otherwise there would be a
characteristic scale where the dynamics would no longer be
dominated by memory, and where the avalanche
hierarchy would break down.  The propagator at time $t$ gets
contributions from {\it all} previous times $(t' \leq t)$ weighted by
$V(t-t')$, which is scale free.  Since we consider only the behavior
in the tail of the distribution, $r^D/t\gg 1$, an equation linear in
$G$ is sufficient. A complete equation for the propagator, such that
probability is conserved, would be far more complicated and would most
likely be nonlinear [as in Eq.~(\ref{envelope}) for the explicit
example below].

We checked Eq.~(\ref{n4})
numerically for the two dimensional Bak-Sneppen model using
$\lambda_c=0.328855$ and $D=2.92$ as determined in
Ref.~\cite{scaling}.  Based on previous numerical studies of
the multi-trait model \cite{BoPa2}, we assume  that $G(r,t)$ has
the scaling form \cite{comment1}
\begin{equation}
G(r,t)\sim t^{\delta-1} g\left({r^D\over t}\right)\quad
\left(r^D/t\gg 1\right).
\label{scal}
\end{equation}
At each update $t$ we determine the location $r$ of the minimal random
number in a surviving avalanche that started at $t=0,~r=0$. We bin
counts as a function of $\log_2 t$ and $\log_2 (r^D/t)$ for avalanches
of duration $t<10^9$; see Fig.~\ref{scaling}.  The data indicates
that the counting rates for each bin rise with $t^\delta,~\delta\approx
0.7,$ for each value of
$r^D/t$. We then ``collapse'' the data to determine $g(x)\sim
t^{1-\delta}G(r,t)$ with the scaling variable $x=r^D/t$ by averaging the
data for each $x$.  We attempted
to numerically determine the asymptotic tail of the function $g$ for
large values of $x$ by assuming in accordance with Eq.~(\ref{n4}) that
\begin{eqnarray}
g(x)\sim \exp\left[-C x^{1\over D'-1}\right],\quad x\gg 1.
\label{gx}
\end{eqnarray}
Since the estimate of $D'$ depends on taking two logarithms of
$g$ to extract the asymptotic behavior for large $x$, it is
not very accurate.  
{}From the extrapolation in Fig.~\ref{extrap} we estimate that
$1/(D'-1)=0.55\pm 0.10$, consistent with the expected result 
of $0.52$ obtained for $D=2.92$ \cite{scaling}. Note that the apparent
convergence of the extrapolation justifies Eq.~(\ref{gx}).

 Eqs.~(\ref{e1},\ref{e2}) can be derived
by considering a similar model
which is analytically tractable, the multi-trait evolution model.  This
model has all of the characteristic punctuated equilibrium behavior of
SOC  including dimension-dependent critical
exponents.  Now,  each site on the lattice has a collection of $M$ random
numbers.  The smallest among all $M\times L^d$ is chosen.  That
particular random number and one random number out of $M$ at each of the $2d$
neighboring sites are replaced by new random numbers independently
drawn from a flat distribution in the unit interval.  When $M=1$, the
Bak-Sneppen model is recovered.  When $M\rightarrow \infty$ the
equation of motion can be obtained explicitly and solved.  In
what follows we choose $M=\infty$, $d=1$, and modify the model
slightly by replacing the minimal random number with the value
1. Then, $\lambda_{\rm c}=1/2$ exactly \cite{BoPa}.

Let $F(r,t)$ be the probability for a $\lambda_{\rm c}$ avalanche 
to survive precisely $t$
steps and to have affected a particular site of distance $r$ from its
origin. Conceptually, the quantity $F(r,t)$ may roughly
correspond to an envelope function of the propagator $G(r,t)$. Due to
the lack of any scale in the model, it is plausible that the
asymptotic behavior of $G$ and $F$ is identical, as comparison with
numerical calculations suggests \cite{BoPa2}.

To simplify the derivation of an equation for $F$, we consider $P(r,t)$, 
the probability  that the $\lambda_{\rm c}$ avalanche dies precisely 
after $t$ updates and does  {\it not} affect a particular site $r$ away 
from the origin of the avalanche. Clearly, 
\begin{equation}
P(r,t)= P(r=\infty, t) - F(r,t) ,
\label{deffrs}
\end{equation}
where $P(r=\infty,t)=P(t)$.  Since no avalanche of finite duration
spreads to an infinitely distant site, $P(t)$ is  the probability
for an avalanche to end exactly after $t$ updates, or its lifetime
distribution. In Ref. \cite{BoPa} it was shown that in this model
$P(t)$ is given by the return time distribution for a random walk,
\begin{equation}
P(t)= {\Gamma(t+1/2) \over \Gamma(1/2)\Gamma(t+2)} \ \ \sim
\pi^{-1/2}t^{-3/2}\quad  {\mbox {\rm for }  } \quad t\gg 1 \quad . 
\label{lifetime}
\end{equation}
 Since we consider the avalanche to
start with a single active value at $r=0$ and $t=0$, 
$P(r=0,t)\equiv 0$ for all $t\geq 0$, and $P(r,t=0)\equiv 0$ for all
$r$.  The remaining properties of a $\lambda_{\rm c}$ avalanche can be
deduced from the properties of avalanches that ensue after the first
update. It will terminate with probability $(1-\lambda_{\rm c})^2=1/4$
after the first update.
Thus, $P(r,t=1)\equiv 1/4$. For avalanches surviving
until $t\geq 2$ we find for $r\geq 1$
\begin{eqnarray}
P(r,t)&=&{1\over 4}\left[P(r-1,t-1) +P(r+1,t-1)\right]\cr
\noalign{\medskip}
&&\quad+{1\over 4}\sum_{t'=0}^{t-1} P(r-1,t')  P(r+1,t-1-t')
\label{y1}
\end{eqnarray}
expressing that the first update may create exactly one new active 
barrier with probability $\lambda_{\rm c}(1-\lambda_{\rm c})=1/4$ either 
to the left or to the right of the origin,
and also that the first update may create two new  active
barriers with probability $\lambda_{\rm c}^2=1/4$ to the left and the right  
of the origin. Then, the properties of the original avalanche of duration $t$
are related to the properties of all combinations of two avalanches of
combined duration $t-1$. Both of these avalanches evolve  in a  {\sl
statistically independent}  manner for $M=\infty$.  
 For any such pair, the probability to
not affect the chosen site of distance $r$ from the origin is given
simply by the product of the probabilities for the two ensuing
avalanches to not affect a chosen site of distance $r-1$ or $r+1$,
respectively.  In this respect, the theory is similar to the
mode-coupling theory used in Refs.\cite{dgl,cule},
except that in our case the theory is exact.
Note that setting $r=\infty$ gives a nonlinear equation
for $P(t)$ whose solution is Eq. (\ref{lifetime}).

The equation governing the envelope function $F(r,t)$ is
obtained by inserting Eq.~(\ref{deffrs}) into Eq.~(\ref{y1}). 
This gives
$F(0,t)\equiv P(t)$, $F(r,0)\equiv 0$, $F(r\geq 1, t=1)=0$,
%COMMENT i changed this again
and for $t\geq 1$, $r\geq 1$,
\begin{eqnarray}
F(r,t+1)&=& {1\over 4}\left[F(r-1,t) +F(r+1,t)\right]\cr
\noalign{\medskip}
&&\quad+{1\over 4}\sum_{t'=0}^{t} P(t-t')\left[F(r-1,t')
+F(r+1,t')\right]\cr
\noalign{\medskip}
&&\quad-{1\over 4}\sum_{t'=0}^{t} F(r-1,t') F(r+1,t-t').
\label{envelope}
\end{eqnarray}

One can show that $F(r,t)\to 0$ for $r^D\gg t\gg 1,~D=4,$ sufficiently
fast such that the nonlinear term in Eq.~(\ref{envelope}) can be
neglected. We can take the continuum limit and obtain \cite{sigma}
\begin{equation} 
{\partial F(r,t)\over \partial t}= {1\over 2}\nabla_r^2 F(r,t) +
{1\over 2}\int_0^t V(t-t')F(r,t')dt' , 
\label{nonlocaleq} 
\end{equation} 
with a nonlocal kernel $V(t)=P(t)-2\delta(t)$.  Aside from trivial
constants, Eq.~(\ref{nonlocaleq}) is identical to Eq.~(\ref{e1}).  In
the absence of the history-dependent potential, the system would be
purely diffusive with a Gaussian tail $F \sim e^{-r^2/2t}$. In its
presence, the probability to have reached a site at distance $r$ at
time $t$ gets contributions from avalanches that reached $r$ at
earlier times $t'<t$ weighted according to $P(t-t')$ which has a
power-law tail.  Avalanches contributing to $F(r,t)$ consist of
sub-avalanches, one of which reaches $r$ in time $t'$ while the
combined duration of the other is $t-t'$.  The avalanche hierarchy
gives a hierarchy of time scales changing the relaxation dynamics to
be non-Gaussian.

Using a Laplace transform, ${\tilde G}(r,y)=\int_0^\infty dt~e^{-
yt}G(r,t)$, Eq.~(\ref{e1}) turns into an ordinary second-order
differential equation in $r$,
\begin{equation}
\nabla_r^2 {\tilde G}(r,y)\sim\left[2y-{\tilde V}(y)\right] {\tilde
G}(r,y),
\label{Laplace}
\end{equation}
where ${\tilde V}(y)$ is the Laplace transform of $V(t)$;
${\tilde V}(y)\sim y^{\alpha-1}$ for small
$y$.  In our particular model where $V(t)\sim P(t)\sim t^{-3/2}$, 
$\alpha=3/2$. We require that $\alpha>1$ so that $V$ can be
normalized. On the other hand, we require $\alpha<2$, so that ${\tilde
V}(y)\gg y$ for $y\to 0$ (i.e. for $t\rightarrow \infty$) in
Eq.~(\ref{Laplace}).  Then in the limit of large times, the effect of
the history-dependent potential dominates over the time derivative in
Eq.~(\ref{e1}), signaling the deviation from simple diffusive
behavior.

The solution of Eq. (11) which decreases for large r is
\begin{equation}
{\tilde G}(r,y)\sim \exp\left[-A r y^{{\alpha-1}\over 2}\right],
\label{expo}
\end{equation}
where $A$ is a constant, and we assume that any prefactor in Eq.~(\ref{expo})
is a sufficiently well-behaved function near $y=0$.

The inverse Laplace transform yields a contour integral with a contour
extending just to the right ($\eta>0$) of the imaginary $y$-axis:
\begin{equation}
G(r,t)\sim\int_{-i\infty+\eta}^{i\infty+\eta}{dy\over 2\pi i} 
\exp\left[yt-A r y^{{\alpha-1}\over 2}\right].
\label{saddle}
\end{equation}
{}For $r/t^{(\alpha-1)/2}\to\infty$ the integral in Eq.~(\ref{saddle})
has a nontrivial saddle point such that $G(r,t)$ is exponentially cut
off, determining the avalanche dimension $D={2\over
\alpha-1}$. With $1<\alpha<2$ we have $D>2$, i. e. the avalanche
dynamics as described by Eq.~(\ref{e1}) always evolves in a
subdiffusive manner. We note that almost all isotropic, extremal SOC
models have been found to have $D>2$ \cite{scaling}.  The notable
exception is the one-dimensional interface depinning model proposed by
Sneppen \cite{sneppen}, which therefore cannot be described by this
theory.  In our particular case we find $D=4$ from $\alpha=3/2$.  In
the limit $r^D/t\to\infty$, we can perform a steepest-descent analysis
of the integral \cite{BeOr} and obtain the non-Gaussian tail given in
Eq.~(\ref{n4}).  In the same way, the complete leading asymptotic behavior for
$F$ in Eq. (9) can be calculated explicitly to give \cite{BoPa2}
\begin{equation} 
F(r,t) \sim \sqrt{24\over\pi} t^{-{3\over 2}}
\left({r^4\over t}\right)^{1\over 3} e^{-{3\over 4} \left({r^4\over
t}\right)^{1\over 3}} \qquad\left(r^4 \gg t \gg 1\right) \quad .
\label{nongaussian} 
\end{equation} 

Note that previous continuum theories for SOC, in particular the
theory of singular diffusion \cite{carlson} by Carlson and others,
specifically apply to systems with a conserved quantity; whereas in the
cases considered here, there is no conservation law.  This does not rule
out the possibility that equations similar to Eq. (1-3) could describe
avalanches in, for instance, sand pile models, which involve
transport of a conserved quantity.  Critical exponents for
avalanches in a limited local sand pile model \cite{limited} have been
related to the singularity in the diffusion equation \cite{maslov}
where the avalanches are averaged in a coarse graining procedure.
This topic is currently under investigation.

It has previously been speculated that glassy dynamics may take place near a
(self-organized) critical point \cite{tangbak}, with mixed results
\cite{jacobs}.  Here we find an anomalous tail in the probability
distribution for the activity in SOC which is characteristic of
glassy systems.  For example, the directed polymer in a random media,
which contains many features of frustrated random systems, also has a
non-Gaussian tail for $G(x,t)$ \cite{halpin}.  Our problem, however,
is inherently dynamical with no quenched disorder.  From a different
viewpoint, Stein and Newman \cite{stein} have put forward a picture of
slow dynamics on a high dimensional rugged fitness landscape based on
an invasion percolation or SOC picture.  Finally, the mode-coupling
equations for glasses \cite{dgl} such as those derived for the
non-ergodic dynamics of the random phase Sine-Gordon model \cite{cule}
are similar to what we find for avalanches in self-organized
criticality.  It would be interesting to see if the mode coupling
equations in glasses could be interpreted in terms of avalanches
sweeping through a critical system.

\section{Acknowledgements}

We thank Y. Shapir, J.P. Bouchaud, J.M. Carlson, and P. Bak for
interesting and helpful remarks.  This work was supported by the
U. S. Department of Energy under Contract No. DE-AC02-76-CH00016 and
DE-FE02-95ER40923.  MP thanks the U.S. Department of Energy
Distinguished Postdoctoral Fellowship Program for partial financial
support, and the Santa Fe Institute where part of this work was
completed.

\figure{\label{scaling}
Counting rates for the activity plotted as a function of time $t$
for the two dimensional Bak-Sneppen model. Each
curve is labeled in descending order by $i=\lceil\log_2 x\rceil$ where
$x=r^D/t$ is the scaling variable [see Eq.~(\ref{gx})].}
\epsfxsize=350pt
\epsfysize=450pt
\epsffile{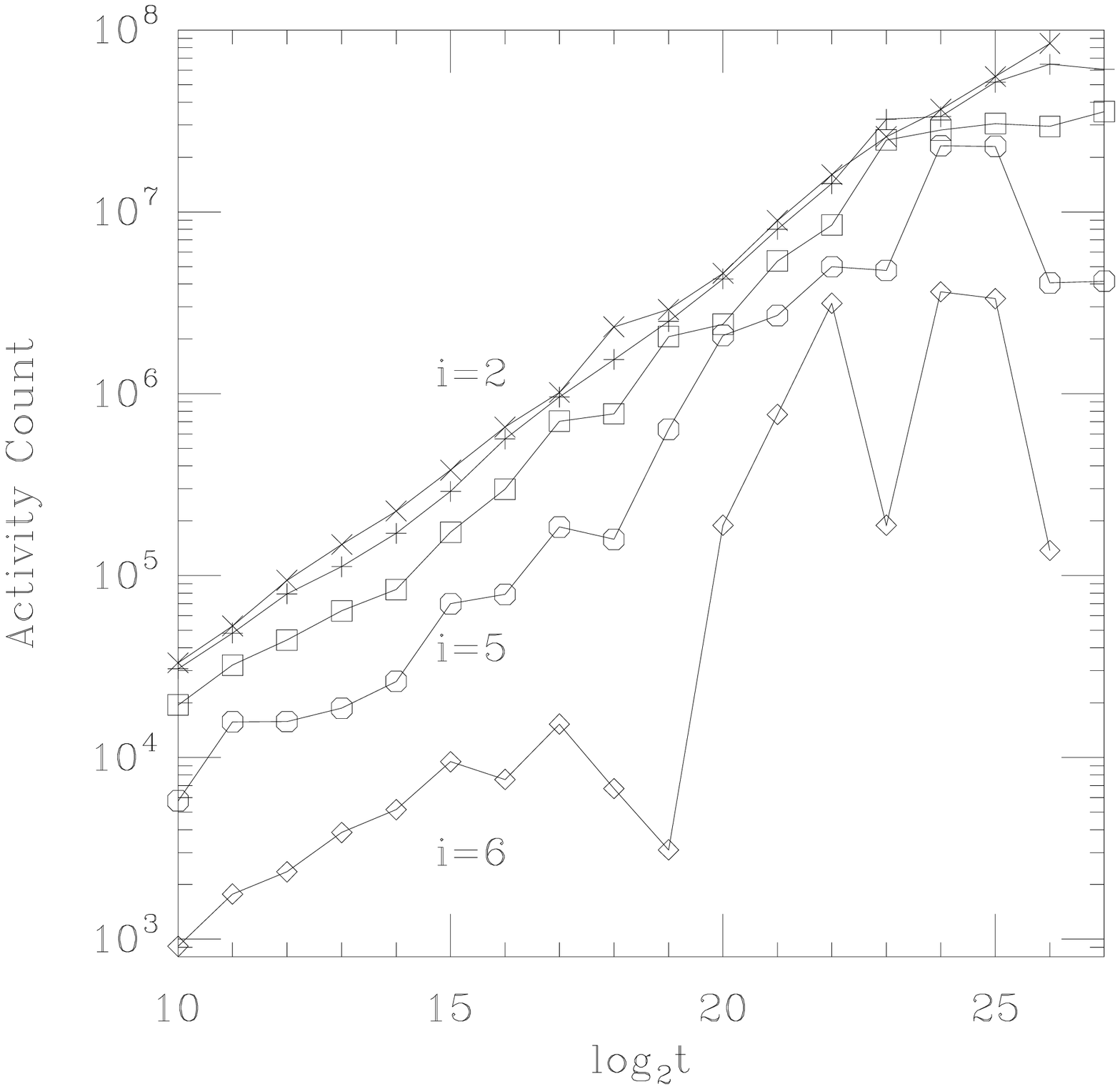}
\newpage

\figure{\label{extrap}
Extrapolation of the data in Fig.~\ref{scaling} according to Eq.~(\ref{gx}).
We plot for each $i=\lceil\log_2 x\rceil$ the projected value of
$1/(D'-1)\sim \ln\left[-\ln g(x)\right]/\ln x$ as a function of $1/\ln
x$ and determine the value for $1/(D'-1)=0.55\pm 0.10$ from the 
extrapolation of this sequence for $x\to\infty$, i. e. at $1/\ln x=0$.
Error bars originate from the averaging of the data in Fig.~\ref{scaling}
for each $i$.}
\epsfxsize=350pt
\epsfysize=450pt
\epsffile{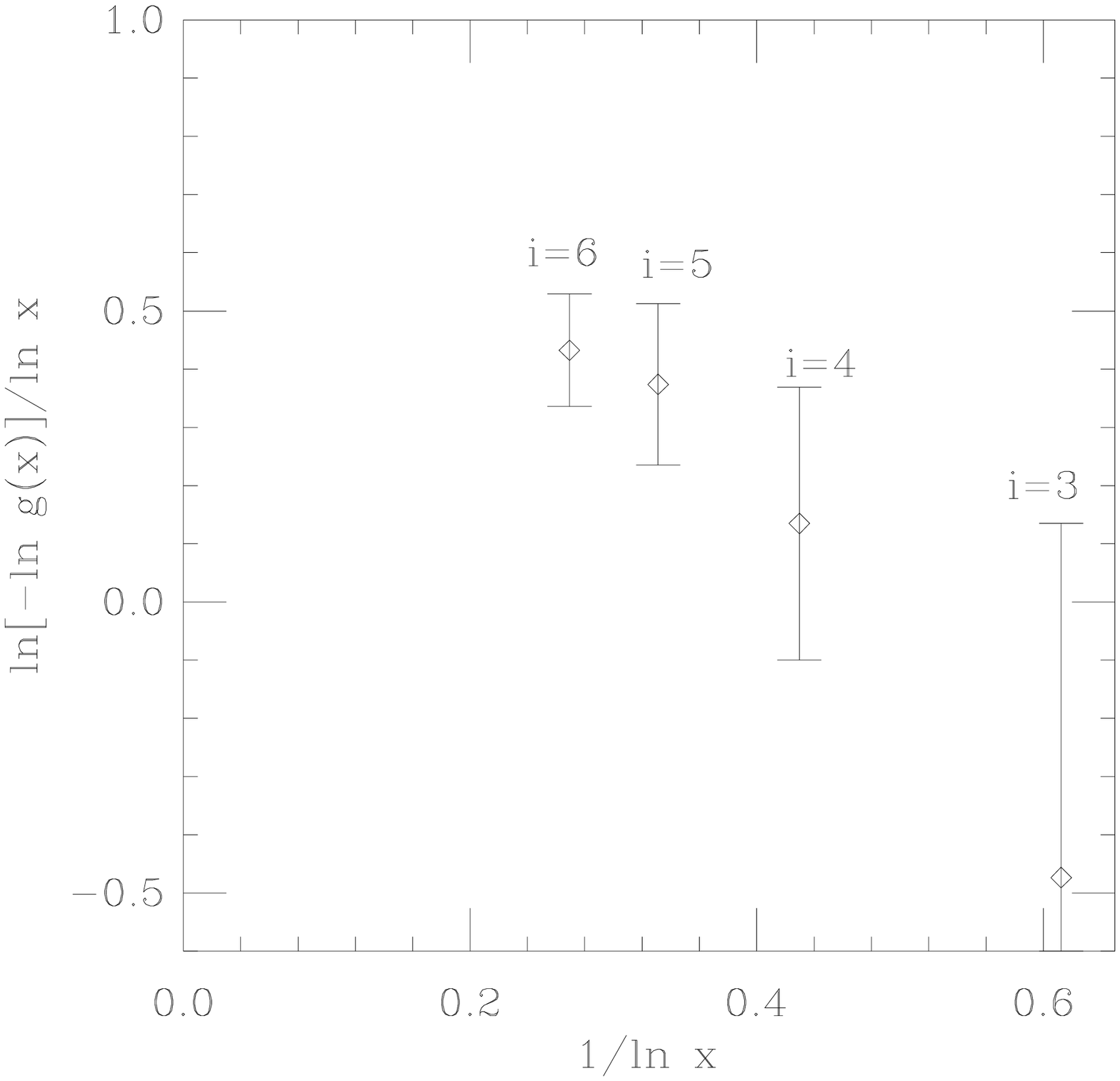}

\begin{references}
\bibitem{sigma}
The origin of the factor preceding the Laplacian is not trivial. This
term in Eq.~(\ref{nonlocaleq}) reads in more detail ${1\over 4} \int_0^t
dt' \sigma(t-t') \nabla_r^2 F(r,t')$ with $\sigma(t)=P(t) + 2\delta(t)$
as a nonlocal diffusion constant. For large $t$ the effect of $\sigma$
becomes local, leading to the given equation.
\end{references}
\end{document}